\documentclass[preprint,12pt]{elsarticle}
\pdfoutput=1

\usepackage{amsmath}
\renewcommand{\vec}[1]{\mathbf{#1}}

\usepackage{amssymb}

\usepackage[colorlinks,
           linkcolor=orange,
           anchorcolor=blue,
           citecolor=blue
           ]{hyperref} 

\usepackage{tabularx}
\usepackage{gensymb}

\biboptions{numbers,sort&compress}

\journal{Computer Physics Communications}

\begin{document}

\begin{frontmatter}



\title{FourPhonon: An extension module to ShengBTE for computing four-phonon scattering rates and thermal conductivity}

\author[label1]{Zherui Han}
\author[label2]{Xiaolong Yang}
\author[label2]{Wu Li}

\author[label3]{Tianli Feng}
\author[label1]{Xiulin Ruan\corref{author}}
\address[label1]{School of Mechanical Engineering and the Birck Nanotechnology Center, Purdue University, West Lafayette, Indiana 47907-2088, USA}
\address[label2]{Institute for Advanced Study, Shenzhen University, Nanhai Avenue 3688, Shenzhen 518060, China}
\address[label3]{Department of Mechanical Engineering, University of Utah, Salt Lake City, Utah 84112, USA}
\cortext[author] {Corresponding author.\\\textit{E-mail address:} ruan@purdue.edu}

\begin{abstract}

\texttt{FourPhonon} is a computational package that can calculate four-phonon scattering rates in crystals. It is built within \texttt{ShengBTE} framework, which is a well-recognized lattice thermal conductivity solver based on Boltzmann transport equation. An adaptive energy broadening scheme is implemented for the calculation of four-phonon scattering rates. In analogy with \verb$thirdorder.py$ in \texttt{ShengBTE}, we also provide a separate python script, \verb$Fourthorder.py$, to calculate fourth-order interatomic force-constants. The extension module preserves all the nice features of the well-recognized lattice thermal conductivity solver \texttt{ShengBTE}, including good parallelism and straightforward workflow. In this paper, we discuss the general theory, program design, and example calculations on Si, BAs and $\mathrm{LiCoO_2}$.

\end{abstract}




\end{frontmatter}


{\bf PROGRAM SUMMARY}

\begin{small}
\noindent
{\em Program Title:}  \texttt{FourPhonon}                         \\
{\em CPC Library link to program files:} (to be added by Technical Editor) \\
{\em Developer's repository link:} \url{https://github.com/FourPhonon} \\
{\em Code Ocean capsule:} (to be added by Technical Editor)\\
{\em Licensing provisions:} GNU General Public License, version 3  \\
{\em Programming language:}     Fortran 90, MPI                              \\
{\em Nature of problem:}\\
Calculation of lattice thermal conductivity and related quantities, determination of both three-phonon and four-phonon scattering rates \\
{\em Solution method:}\\
Four-phonon scattering rates at RTA level, adaptive broadening scheme\\
{\em Additional comments including restrictions and unusual features:}\\
 For a productive run, one needs to use HPC facilities and it takes several hours to several days to finish calculations
   \\
\end{small}

\section{Introduction}

Phonons, quantum mechanical description of atomic vibrations in crystals, are the main heat carriers in most insulators, semiconductors and some semimetals. Phonon properties and lattice thermal conductivity are largely determined by phonon-phonon interactions. Starting from 1929 \cite{peierls1929kinetischen} when Peierls proposed the first formulation to calculate thermal conductivity using Boltzmann transport equation (BTE), subsequent models \cite{callaway1959model,ziman1960electronsphonons,Maradudin1962AnharmonicCrystal,omini1995iterative} with fewer fitting parameters have emerged owing to the advancement of both theory and computational power. About a decade ago, Broido et al. \cite{broido2007intrinsic} combined first-principles calculations and phonon Boltzmann transport equation, and enabled first-principles prediction of thermal conductivity. This method has gained great success in predicting thermal properties, and not until recently the common practice was to consider phonon-phonon interactions up to the lowest order, i.e., three-phonon scattering \cite{Esfarjani2011Si,lindsay2013BAs,seko2015prediction,review2018mtp,JAPreviewMcGaughey}. Among various computational tools for calculating the lattice thermal conductivity, \verb$ShengBTE$ \cite{2014shengbte} is a well-developed and mostly used software package. 

Phonon-phonon interactions stem from the intrinsic lattice anharmonicity of materials. While three-phonon scattering was assumed to be adequate in describing anharmonicity \cite{broido2007intrinsic,Esfarjani2011Si,lindsay2013BAs}, Feng and Ruan developed a rigorous four-phonon scattering formalism \cite{4phPRB2016}, and have revealed that higher-order anharmonicity can play a significant role. In 2017, Feng , Lindsay, and Ruan predicted that four-phonon scattering is surprisingly strong in BAs  \cite{prb2017RC}, a material predicted to possess higher thermal conductivity than diamond by three-phonon scattering theory \cite{lindsay2013BAs}. In 2018, three independent experiments \cite{scienceHu2018,scienceChen2018,scienceCahill2018} confirmed the significance of four-phonon scattering in BAs. Subsequent studies pursuing four-phonon effect have proved its importance in a broader range of systems \cite{4ph-graphene,XiaYi2018,Broido4phPRB2018,linewidth4ph2020,Briodo4phPRX2020,kundu2021ultrahigh}, including insulators, semiconductors and semimetals with topics covering thermal conductivity predictions, radiative properties, and phonon linewidths. Future research into material properties at high temperatures, optical phonon energy decay, infrared properties or other interesting topics may require to include four-phonon scattering as well. A detailed discussion can be found in Ref.\cite{bookchapter4ph}. Despite the evident importance of four-phonon scattering calculations, they have remained largely inaccessible to the thermal transport community due to the complex form and large computational load. To the best of our knowledge, there is not an open-source software available yet to perform four-phonon calculations and the newly developed tool presented in this paper should be beneficial.

Here, we present a computational package for four-phonon scattering calculations, \verb$FourPhonon$. We develop this tool as an extension module to \verb$ShengBTE$, which has a good extendibility. With a subroutine \verb$Fourthorder.py$ to calculate fourth-order interatomic force-constants(IFCs) through first-principles, \verb$FourPhonon$ is capable of computing four-phonon phase space, scattering rates from both normal and Umklapp processes, and lattice thermal conductivity. Its operation is fully compatible with the current \verb$ShengBTE$ package. 

The remainder of this paper is organized as follows. Section 2 presents the theoretical background and computational details of the \verb$FourPhonon$ package. In Section 3, we show three case studies on representative materials: Si, BAs and $\mathrm{LiCoO_2}$. We also document the input and output files of our software in \ref{appendix.files}.
\section{Methodology and computational details}
\subsection{\texttt{ShengBTE} + \texttt{FourPhonon} workflow}
\begin{figure}
\centerline{\includegraphics[width=6in]{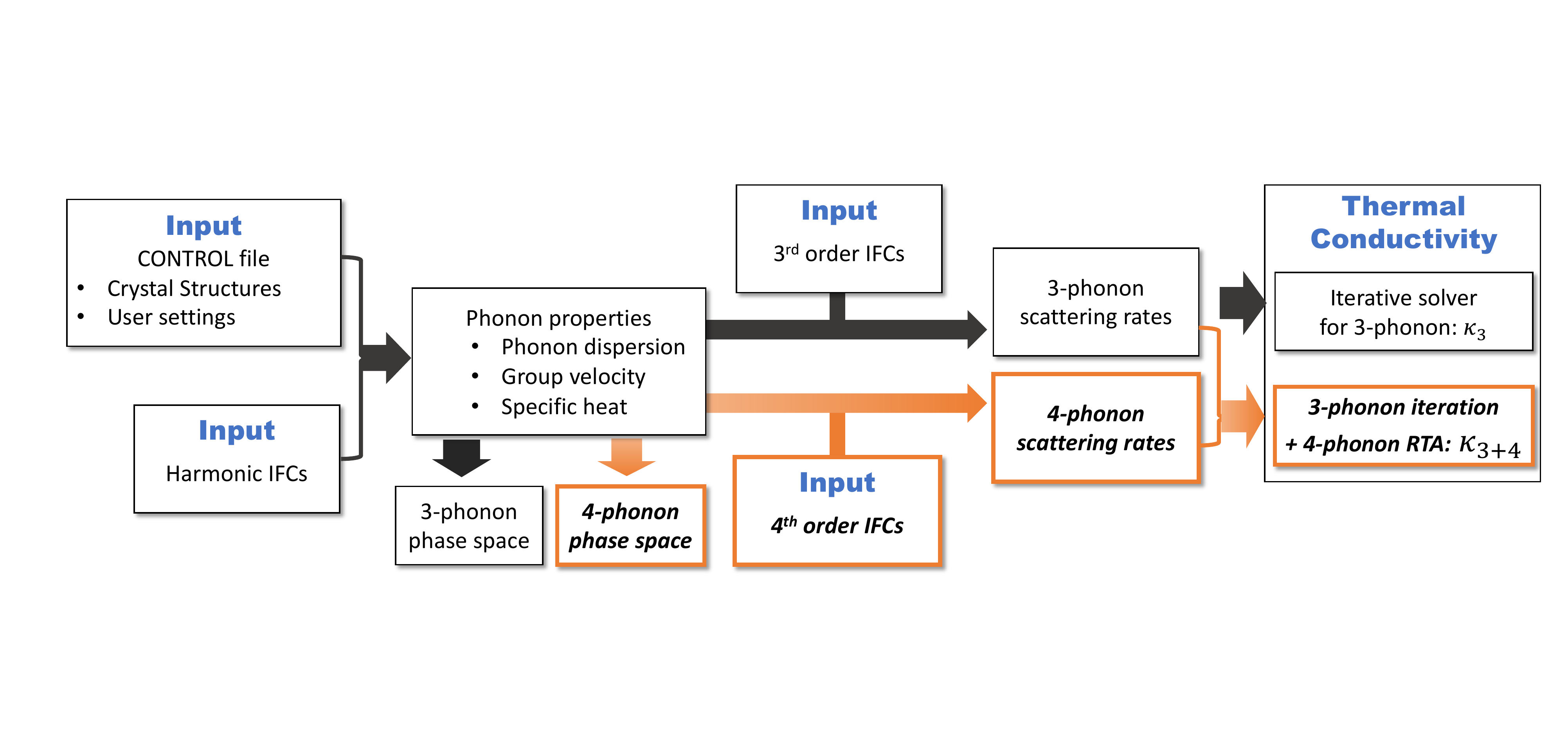}}
\caption{Modified workflow of \texttt{ShengBTE} + \texttt{FourPhonon}. Orange parts represent added modules by \texttt{FourPhonon}.}
\label{workflow}
\end{figure}

\verb$FourPhonon$ is built within \verb$ShengBTE$ framework and its execution is fully compatible with \verb$ShengBTE$. Before explaining \verb$FourPhonon$, we briefly review the basic \verb$ShengBTE$ workflow \cite{2014shengbte}. It uses second-order and third-order force constants as inputs, retrieves phonon properties from harmonic calculations, estimates the available three-phonon scattering phase space, and then calculates the three-phonon scattering rates and lattice thermal conductivity. Besides the $2^{nd}$- and $3^{rd}$-order force constants files, users need to provide another input file, named \verb$CONTROL$, which contains all the user-specified settings and parameters, including crystal structural information, temperature, $\mathbf{q}$-mesh, broadening factor, etc. On the basis of this workflow, \verb$FourPhonon$ requires an additional input file with fourth-order force constants. Flag is also added to the \verb$CONTROL$ file to enable \verb$FourPhonon$ utilities, namely \verb$four_phonon$. When combined with other inherent flags or settings, users can perform various tasks, gaining insight into four-phonon scatterings. We now illustrate the modified workflow in more details. A workflow chart is presented in Fig. \ref{workflow}.

Due to the great computational cost of four-phonon scattering in most materials, one may need to estimate first how significant four-phonon scattering is compared to three-phonon scattering in a certain system. One can use the available three-phonon and four-phonon phase spaces as a criterion. To calculate these values, users can turn on the flags \verb$onlyharmonic$ and \verb$four_phonon$, which will count the number of four-phonon scattering processes so that one can have a quick estimation of the computational demand when going forward. 

Another particular problem is one may need to determine whether to solve the BTE exactly with an iterative scheme beyond single mode relaxation time approximations (SMRTA, or RTA) to get converged thermal conductivity. Some theoretical works have indicated that this is closely related to the relative strength of phonon normal/Umklapp processes \cite{lindsay2010ZA,4ph-graphene}. Since the iterative scheme would require huge computational resources (both processors and memory), users can first perform calculations in SMRTA scheme by turning on \verb$four_phonon$ flag and disable \verb$convergence$ flag. 
SMRTA calculations usually take much less time to finish than iterative scheme and would output scattering rates from normal/Umklapp scattering events separately. If Umklapp processes are dominant, then the accuracy in thermal conductivity may be acceptable without further utilizing iterative scheme. If normal processes are relatively strong, one can apply the iterative scheme to BTE up to three-phonon scattering by turning on \verb$convergence$ flag, with four-phonon scattering rates treated at the RTA level. Presently, this package is not able to involve the four-phonon scattering phase space in the iteration scheme due to the exceptionally high memory demands. Further optimization could be made to include this functionality in the future.

\subsection{Implemented four-phonon formalism}
The linearized phonon Boltzmann transport equation with three-phonon scattering has been extensively studied in the literature, and the formalism on four-phonon scattering has also been well documented in several preceding papers \cite{4phPRB2016,prb2017RC,4ph-graphene}. Here, we only introduce a unified formalism implemented in our program and make all the corresponding notations consistent with the one used in \texttt{ShengBTE} \cite{2014shengbte}. For a certain mode $\lambda$ , the linearized BTE is expressed as:
\begin{equation}
    \mathbf{F_\lambda}=\tau^0_\lambda(\mathbf{v_\lambda}+\mathbf{\Delta_\lambda})
    \label{BTE}
\end{equation}
where $\mathbf{F_\lambda}$ is the linear coefficient in nonequilibrium phonon distribution function and $\tau^0_\lambda$ is the relaxation time for mode $\lambda$ under SMRTA. $\mathbf{\Delta_\lambda}$ works for iterative scheme and is a quantity showing the phonon population deviation from the SMRTA scheme. With the inclusion of four-phonon scattering, $\mathbf{\Delta_\lambda}$ and $\tau^0_\lambda$ (0 represents zeroth-order in iterations) are computed as:
\begin{equation}
    \begin{aligned}
    \mathbf{\Delta_\lambda}=&\frac{1}{N}\displaystyle\sum^{(+)}_{\lambda'\lambda''}\Gamma^{(+)}_{\lambda\lambda'\lambda''}(\xi_{\lambda\lambda''}\mathbf{F_{\lambda''}}-\xi_{\lambda\lambda'}\mathbf{F_{\lambda'}})+\frac{1}{N}\displaystyle\sum^{(-)}_{\lambda'\lambda''}\frac{1}{2}\Gamma^{(-)}_{\lambda\lambda'\lambda''}(\xi_{\lambda\lambda''}\mathbf{F_{\lambda''}}+\xi_{\lambda\lambda'}\mathbf{F_{\lambda'}})\\
    &\left.
   \begin{array}{cc}
     &+\frac{1}{N}\displaystyle\sum^{(++)}_{\lambda'\lambda''\lambda'''}\frac{1}{2}\Gamma^{(++)}_{\lambda\lambda'\lambda''\lambda'''}(\xi_{\lambda\lambda'''}\mathbf{F_{\lambda'''}}-\xi_{\lambda\lambda'}\mathbf{F_{\lambda'}}-\xi_{\lambda\lambda''}\mathbf{F_{\lambda''}})
    \\&+\frac{1}{N}\displaystyle\sum^{(+-)}_{\lambda'\lambda''\lambda'''}\frac{1}{2}\Gamma^{(+-)}_{\lambda\lambda'\lambda''\lambda'''}(\xi_{\lambda\lambda'''}\mathbf{F_{\lambda'''}}-\xi_{\lambda\lambda'}\mathbf{F_{\lambda'}}+\xi_{\lambda\lambda''}\mathbf{F_{\lambda''}})
    \\&+\frac{1}{N}\displaystyle\sum^{(--)}_{\lambda'\lambda''\lambda'''}\frac{1}{6}\Gamma^{(--)}_{\lambda\lambda'\lambda''\lambda'''}(\xi_{\lambda\lambda'''}\mathbf{F_{\lambda'''}}+\xi_{\lambda\lambda'}\mathbf{F_{\lambda'}}+\xi_{\lambda\lambda''}\mathbf{F_{\lambda''}})
   \end{array}
   \right\rbrace {\text{four-phonon\ terms}}\\
    &+\frac{1}{N}\displaystyle\sum^{(iso)}_{\lambda'}\Gamma^{(iso)}_{\lambda\lambda'}\xi_{\lambda\lambda'}\mathbf{F_{\lambda'}},
    \label{delta}
    \end{aligned}
\end{equation}

\begin{equation}
    \begin{aligned}
    \frac{1}{\tau^0_\lambda}=&\frac{1}{N}\left(\displaystyle\sum^{(+)}_{\lambda'\lambda''}\Gamma^{(+)}_{\lambda\lambda'\lambda''}+\displaystyle\sum^{(-)}_{\lambda'\lambda''}\frac{1}{2}\Gamma^{(-)}_{\lambda\lambda'\lambda''}\right)+\frac{1}{N}\displaystyle\sum^{(iso)}_{\lambda'}\Gamma^{(iso)}_{\lambda\lambda'}
    \\&+\underbrace{\frac{1}{N}\left(\displaystyle\sum^{(++)}_{\lambda'\lambda''\lambda'''}\frac{1}{2}\Gamma^{(++)}_{\lambda\lambda'\lambda''\lambda'''}+\displaystyle\sum^{(+-)}_{\lambda'\lambda''\lambda'''}\frac{1}{2}\Gamma^{(+-)}_{\lambda\lambda'\lambda''\lambda'''}+\displaystyle\sum^{(--)}_{\lambda'\lambda''\lambda'''}\frac{1}{6}\Gamma^{(--)}_{\lambda\lambda'\lambda''\lambda'''}\right)}_{\text{four-phonon\ scattering\ terms}},
    \label{tao0}
    \end{aligned}
\end{equation}
where $N$ is the total grid of $\Vec{q}$ points.  $\xi_{\lambda\lambda'}=\omega_{\lambda'}/\omega_{\lambda''}$. The superscripts $(\pm)$ or $(\pm\pm)$ represent the three-phonon (3ph) and four-phonon (4ph) processes, namely $\Vec{q}''=\Vec{q}\pm\Vec{q'}+\vec{Q}$ and $\Vec{q}'''=\Vec{q}\pm\Vec{q'}\pm\Vec{q''}+\vec{Q}$, respectively. $\vec{Q}$ is a reciprocal lattice vector with $\vec{Q}=0$ implying normal process. Note that Eq.(\ref{delta}) is the full expression for three- and four-phonon iterative scheme, whereas the current version of \verb$FourPhonon$ does not support such functionality and  excludes terms in the right bracket in Eq.(\ref{delta}). $\Gamma$ denotes the scattering rates for different processes and are computed by scattering probability matrices \cite{maradudin1962scattering,4phPRB2016}:
\begin{equation}
    \begin{aligned}
    &\Gamma^{(+)}_{\lambda\lambda'\lambda''}=\frac{\hbar\pi}{4}\frac{n^0_{\lambda'}-n^0_{\lambda''}}{\omega_\lambda\omega_{\lambda'}\omega_{\lambda''}}|V_{\lambda\lambda'\lambda''}^{(+)}|^2\delta(\omega_\lambda+\omega_{\lambda'} -\omega_{\lambda''})\\
    &\Gamma^{(-)}_{\lambda\lambda'\lambda''}=\frac{\hbar\pi}{4}\frac{n^0_{\lambda'}+n^0_{\lambda''}+1}{\omega_\lambda\omega_{\lambda'}\omega_{\lambda''}}|V_{\lambda\lambda'\lambda''}^{(-)}|^2\delta(\omega_\lambda-\omega_{\lambda'} -\omega_{\lambda''}),
    \label{3phgamma}
    \end{aligned}
\end{equation}
\begin{equation}
    \begin{aligned}
    &\Gamma^{(++)}_{\lambda\lambda'\lambda''\lambda'''}=\frac{\hbar^2\pi}{8N}\frac{(1+n^0_{\lambda'})(1+n^0_{\lambda''})n^0_{\lambda'''}}{n^0_\lambda}|V_{\lambda\lambda'\lambda''\lambda'''}^{(++)}|^2\frac{\delta(\omega_\lambda+\omega_{\lambda'}+\omega_{\lambda''}-\omega_{\lambda'''})}{\omega_\lambda\omega_{\lambda'}\omega_{\lambda''}\omega_{\lambda'''}}\\
    &\Gamma^{(+-)}_{\lambda\lambda'\lambda''\lambda'''}=\frac{\hbar^2\pi}{8N}\frac{(1+n^0_{\lambda'})n^0_{\lambda''}n^0_{\lambda'''}}{n^0_\lambda}|V_{\lambda\lambda'\lambda''\lambda'''}^{(+-)}|^2\frac{\delta(\omega_\lambda+\omega_{\lambda'}-\omega_{\lambda''}-\omega_{\lambda'''})}{\omega_\lambda\omega_{\lambda'}\omega_{\lambda''}\omega_{\lambda'''}}\\
    &\Gamma^{(--)}_{\lambda\lambda'\lambda''\lambda'''}=\frac{\hbar^2\pi}{8N}\frac{n^0_{\lambda'}n^0_{\lambda''}n^0_{\lambda'''}}{n^0_\lambda}|V_{\lambda\lambda'\lambda''\lambda'''}^{(--)}|^2\frac{\delta(\omega_\lambda-\omega_{\lambda'}-\omega_{\lambda''}-\omega_{\lambda'''})}{\omega_\lambda\omega_{\lambda'}\omega_{\lambda''}\omega_{\lambda'''}},
    \label{4phgamma}
    \end{aligned}
\end{equation}
where Eq.(\ref{3phgamma}) is for three-phonon processes and Eq.(\ref{4phgamma}) for the four-phonon processes, with $n^0_\lambda$ being the phonon Bose-Einstein distribution at equilibrium, $\omega_\lambda$ being the phonon frequency for a certain mode $\lambda$. Conservation of energy is enforced by the Dirac delta function $\delta$. In Eq.(\ref{3phgamma}) and (\ref{4phgamma}), the matrix elements $V$ are given by the Fourier transformation of force constants, or transition probability matrices:
\begin{equation}
    \begin{aligned}
    V_{\lambda\lambda'\lambda''}^{(\pm)}=\displaystyle\sum_{ijk}\displaystyle\sum_{\alpha\beta\gamma} \Phi_{ijk}^{\alpha\beta\gamma} \frac{e_{\alpha}^\lambda(i) e_{\beta}^{\pm\lambda'}(j)e_{\gamma }^{-\lambda''}(k)}{\sqrt{\bar M_i \bar M_j\bar M_k}} e^{\pm i\mathbf q'\cdot \mathbf r_j } e^{- i\mathbf q''\cdot \mathbf r_k },
    \label{V3}
    \end{aligned}
\end{equation}
\begin{equation}
    \begin{aligned}
    V_{\lambda\lambda'\lambda''\lambda'''}^{(\pm\pm)}=\displaystyle\sum_{ijkl}\displaystyle\sum_{\alpha\beta\gamma\theta} \Phi_{ijkl}^{\alpha\beta\gamma\theta} \frac{e_{\alpha}^\lambda(i) e_{\beta}^{\pm\lambda'}(j)e_{\gamma }^{\pm\lambda''}(k)e_{\theta}^{-\lambda'''}(l)}{\sqrt{\bar M_i \bar M_j\bar M_k\bar M_l}} e^{\pm i\mathbf q'\cdot \mathbf r_j } e^{\pm i\mathbf q''\cdot \mathbf r_k } e^{-i\mathbf q'''\cdot \mathbf r_l},
    \label{V4}
    \end{aligned}
\end{equation}
where $i,j,k,l$ denote the atomic indices and $\alpha,\beta,\gamma,\theta$ denote the Cartesian dimensions $x,y,or z$. $\Phi_{ijk}^{\alpha \beta \gamma}$ and $\Phi_{ijkl}^{\alpha \beta \gamma \theta}$ are the third-order and fourth-order force constants, respectively. $e_{\alpha}^\lambda(i)$ is the eigenvector component for a mode. $\vec{r}_j$ is the position vector of the unit cell where $j$th atom lies, and $M_j$ is its mass.

The remaining formulae including isotope scatterings and the expression of thermal conductivity are the same as \verb$ShengBTE$, so are not elaborated in this paper.

\subsection{Fourth-order force constants}
The potential energy of  crystals can be written as a Taylor expansion with respect to the atomic displacement based on perturbation theory \cite{maradudin1962scattering}
\begin{equation}
    \begin{aligned}
E=E_0+\frac{1}{2}\displaystyle\sum_{\substack{ij\\\alpha\beta}}\Phi_{ij}^{\alpha\beta}r_i^{\alpha}r_j^{\beta}+\frac{1}{3!}\sum_{\substack{ijk\\\alpha\beta\gamma}}\Phi_{ijk}^{\alpha\beta\gamma}r_i^{\alpha}r_j^{\beta}r_k^{\gamma}+\frac{1}{4!} \sum_{\substack{ijkl\\\alpha\beta\gamma\theta}}\Phi_{ijkl}^{\alpha\beta\gamma\theta}r_i^{\alpha}r_j^{\beta}r_k^{\gamma}r_l^{\theta}+...,
    \label{PE}
    \end{aligned}
\end{equation}
where $r_i^{\alpha}$ is the displacement of atom $i$  along the $\alpha (x, y, z)$ direction, $E_0$ is the potential energy when $r = 0$, and the coefficients $\Phi$ in the $n$-th order term are the corresponding
$n$-th order interatomic force constants (IFCs). Given by the translational symmetry, fourth-order IFCs depend only on the relative coordinates of atom  quartette $(i,j,k,l)$ rather than the absolute coordinates of the four atoms. The IFCs are usually determined by either a finite-difference scheme on the atomic forces \cite{2014shengbte} or fitting the atomic displacement-force relations \cite{2014J.Phys}.

In \verb$Fourthorder.py$ we employ a real-space  finite-difference method to calculate the fourth-order IFCs:
\begin{equation}
    \begin{aligned}
 \Phi_{ijkl}^{\alpha\beta\gamma\theta}=&\frac{\partial^4E}{\partial r_i^{\alpha}\partial r_j^{\beta}\partial r_k^{\gamma}\partial r_l^{\theta}}
\\&\approx \frac{1}{2h}\left[\frac{\partial^3E}{\partial r_j^{\beta}\partial r_k^{\gamma}\partial r_l^{\theta}}(r_i^{\alpha}=h)-\frac{\partial^3E}{\partial r_j^{\beta}\partial r_k^{\gamma}\partial r_l^{\theta}}(r_i^{\alpha}=-h)\right]
\\& \approx \frac{1}{4h^2}\left[\frac{\partial^2E}{\partial r_k^{\gamma}\partial r_l^{\theta}} (r_i^{\alpha}=h, r_j^{\beta}=h)-\frac{\partial^2E}{\partial r_k^{\gamma}\partial r_l^{\theta}} (r_i^{\alpha}=h, r_j^{\beta}=-h) \right.
\\& \left.-\frac{\partial^2E}{\partial r_k^{\gamma}\partial r_l^{\theta}} (r_i^{\alpha}=-h, r_j^{\beta}=h)+\frac{\partial^2E}{\partial r_k^{\gamma}\partial r_l^{\theta}} (r_i^{\alpha}=-h,r_j^{\beta}=-h)\right] 
\\& \approx \frac{1}{8h^3}\left[F_l^{\theta}(r_i^{\alpha}=h, r_j^{\beta}=h,
r_k^{\gamma}=h)-F_l^{\theta}(r_i^{\alpha}=h, r_j^{\beta}=h, r_k^{\gamma}=-h) \right.
\\& \left.-F_l^{\theta}(r_i^{\alpha}=h, r_j^{\beta}=-h, r_k^{\gamma}=h)+F_l^{\theta}(r_i^{\alpha}=h, r_j^{\beta}=-h, r_k^{\gamma}=-h) \right.
\\& \left.-F_l^{\theta}(r_i^{\alpha}=-h, r_j^{\beta}=h, r_k^{\gamma}=h)+F_l^{\theta}(r_i^{\alpha}=-h, r_j^{\beta}=h, r_k^{\gamma}=-h) \right.
\\& \left.+F_l^{\theta}(r_i^{\alpha}=-h, r_j^{\beta}=-h, r_k^{\gamma}=h)-F_l^{\theta}(r_i^{\alpha}=-h, r_j^{\beta}=-h, r_k^{\gamma}=-h)\right],
 \label{fourth-order IFCs}
    \end{aligned}
\end{equation}
where $h$ is a small displacement from the equilibrium position, and $F_l^{\theta}$ is the $\theta$ component of the force on the $l$-th atom. The default value of h is 0.04 times the Bohr radius, which can be changed by editing the header of the \verb$Fourthorder.py$ script. Each element of $\Phi_{ijkl}^{\alpha\beta\gamma\theta}$ requires eight DFT calculations with
different supercell configurations, and there are 81$n^4N^3$  elements a supercell with $N$ unit cells with $n$ basis atoms per unit cell.  A typical fourth-order IFCs calculation requires supercells  containing more than 100 atoms,  meaning that tens of thousands of single DFT simulation would be needed.  Since directly performing these calculations is impractical for most computing infrastructures,  it is crucial to make use of the symmetry
analysis to reduce the computational cost.

We begin by populating the force constant matrices with transposition symmetries, and in the case of fourth-order IFCs this permutation of indices can have 24 sets of equality constraints
\begin{equation}
    \begin{aligned}
 \Phi_{ijkl}^{\alpha\beta\gamma\theta}=\Phi_{ijlk}^{\alpha\beta\theta\gamma}=...
    \label{TS}
    \end{aligned}
\end{equation}
Then,  we apply a space group symmetry operation $\sum_{\alpha}{\bf T}^{\alpha '\alpha}{\bf R}_i^{\alpha}+{\bf b}^{\alpha '}={\bf R}_{{\bf T}_{b(i)}}^{\alpha '}$ to the tensor, where $\bf{T}$ and $\bf{b}$ represent the point-group and translation operators respectively, and ${\bf{T}}_{ b(i)}$ specifies the atom to which the $i$th atom is mapped under the corresponding operation.  The fourth-order IFC tensor must satisfy the following relation:
\begin{equation}
  \Phi_{{\bf T}_{b(i)}; {\bf T}_{b(j)}; {\bf T}_{b(k)}; {\bf T}_{b(l)}}^{\alpha ' \beta ' \gamma ' \theta '}=\sum_{\alpha\beta\gamma\theta}{\bf T}^{\alpha'\alpha}{\bf T}^{\beta'\beta}{\bf T}^{\gamma'\gamma}{\bf T}^{\theta'\theta}\Phi_{ijkl}^{\alpha\beta\gamma\theta}.
    \label{SGSO}
\end{equation}

Each symmetry operation enables to map the four atom indices {$i, j, k, l$} into themselves or to a different set.  The first class of operations enforces $m$ constraints on the set of 81 fourth-order IFCs related to an atomic quartette, and each  constraint can be described by a linear equation. Using Gaussian elimination, the $m$ linear equations can be transformed into the following form:
\begin{equation}
\left(
\begin{matrix}
1 & 0 & *  & 0 & 0 & ... \\
0 & 1 & *  & 0 & 0 & ... \\
0 & 0 & 0  & 1 & 0 & ... \\
0 & 0 & 0 & 0 & 1 & ... \\
\vdots & \vdots & \vdots  & \vdots & \vdots & \vdots \\
\end{matrix}
  \right)\left(
\begin{matrix}
     x_1 \\
     x_2 \\
     x_3 \\
 \vdots \\
    x_{81} \\
\end{matrix}
  \right)=0,
    \label{Linear_equation}
\end{equation}
where asterisks in the columns stand for arbitrary numbers, corresponding to the independent elements among the set of 81 IFCs. The remaining force constants are linear combinations of them.  Likewise with \verb$thirdorder.py$ ,  \verb$Fourthorder.py$ relies on the \verb$spglib library$ for detecting all kinds of crystal symmetries. 

The fourth-order IFC tensor also obey the acoustic sum rules (ASRs)
\begin{equation}
\sum_l\Phi_{ijkl}^{\alpha\beta\gamma\theta}=0.
    \label{ASR}
\end{equation}
Considering transposition symmetries mentioned above, Eq.(\ref{ASR}) must also be valid if the summation is performed over $i$, $j$, or $k$. 

Whether these sums are exactly zero  has a significant effect on the calculated phonon scattering rates at low frequencies  near the ${\bf \Gamma}$  point.  Numerical uncertainties usually lead to small violations in ASRs, leading to significant deviations in calculated scattering rates. To tackle this problem, we adopt the same method with the \verb$thirdorder.py$ \cite{2014shengbte}, namely adding a small compensation to each independent non-zero IFC,  where the compensation is chosen by minimizing the sum of squares of these compensations by introducing a Lagrange multiplier. 

It should also be noted that the fourth-order IFCs convergence with respect to the cutoff interatomic distance has to be checked manually, by examining either the force constants directly or the calculated thermal conductivity values. The computational cost increases significantly with increasing cutoff interatomic distance.  Fortunately, it has been found \cite{prb2017RC,yang2019stronger,tong2020first,linewidth4ph2020,feng2020quantum}  that, for many three-dimensional crystals, including up to second nearest neighbors in the fourth-order IFCs can give a satisfactorily converged values of scattering rates and lattice thermal conductivity.

\subsection{Adaptive broadening for energy conservation}
The delta function $\delta$ for energy conservation is approximated by using the adaptive Gaussian broadening method \cite{Li2012PRBbroadening,Broaden2007}. The expression for four-phonon processes is similar to the three-phonon case. Energy spacing level is described by taking the derivative of energy difference respect to the third phonon involved $\frac{\partial \Delta\omega}{\partial \vec{q}''}$. This adaptive broadening then needs the group velocity of a certain mode $\vec{v}_\lambda$ and the spacing of sampling $\vec{q}$ points $|\Delta \Vec{q}''|$. The formula is expressed as:
\begin{align}
    \delta(\omega_\lambda \pm \omega_{\lambda'} \pm\omega_{\lambda''}-\omega_{\lambda'''}) \approx \frac{1}{\sqrt\pi\sigma}e^{-\frac{(\Delta \omega)^2}{\sigma^2}},
\end{align}
while the parameter $\sigma$ for different processes is calculated as:
\begin{align}
    \sigma=
    |\vec{v}_{\lambda''}-\vec{v}_{\lambda'''}||\Delta \Vec{q}''|
\end{align}
The detailed derivations for this formula can be found in \ref{appendix.formulas}.
Similar to \verb$ShengBTE$, users of this program can always adopt a smaller \verb$scalebroad$ to reduce the computational cost by including fewer four-phonon processes and it can often give reasonable results. Our implementation also addresses a technical problem. When $\sigma$ evaluated is very small, the calculated scattering rates may be some abnormal high values. This numerical error is corrected by setting the $\delta$ function as 1.
\section{Examples}

\subsection{Si}
Being the most commonly used semiconductor, silicon has been extensively studied as a benchmark material for theoretical predictions. We use silicon to illustrate the common practice for calculating thermal conductivity with four-phonon scattering.

Force constants are all calculated by first principles using VASP package \cite{vasp}. Using a $6\times6\times6$ supercell, we have considered up to the fifth-nearest neighboring atoms in the third-order IFCs and second-nearest neighboring atoms in the fourth-order IFCs. Calculation of these IFCs is nearly a standard process and is well documented in some helpful tutorials \cite{JAPreviewMcGaughey}. In the calculation of silicon, we have turned on the iterative flag for three-phonon scattering.

\begin{figure}
\centerline{\includegraphics[width=3.5in]{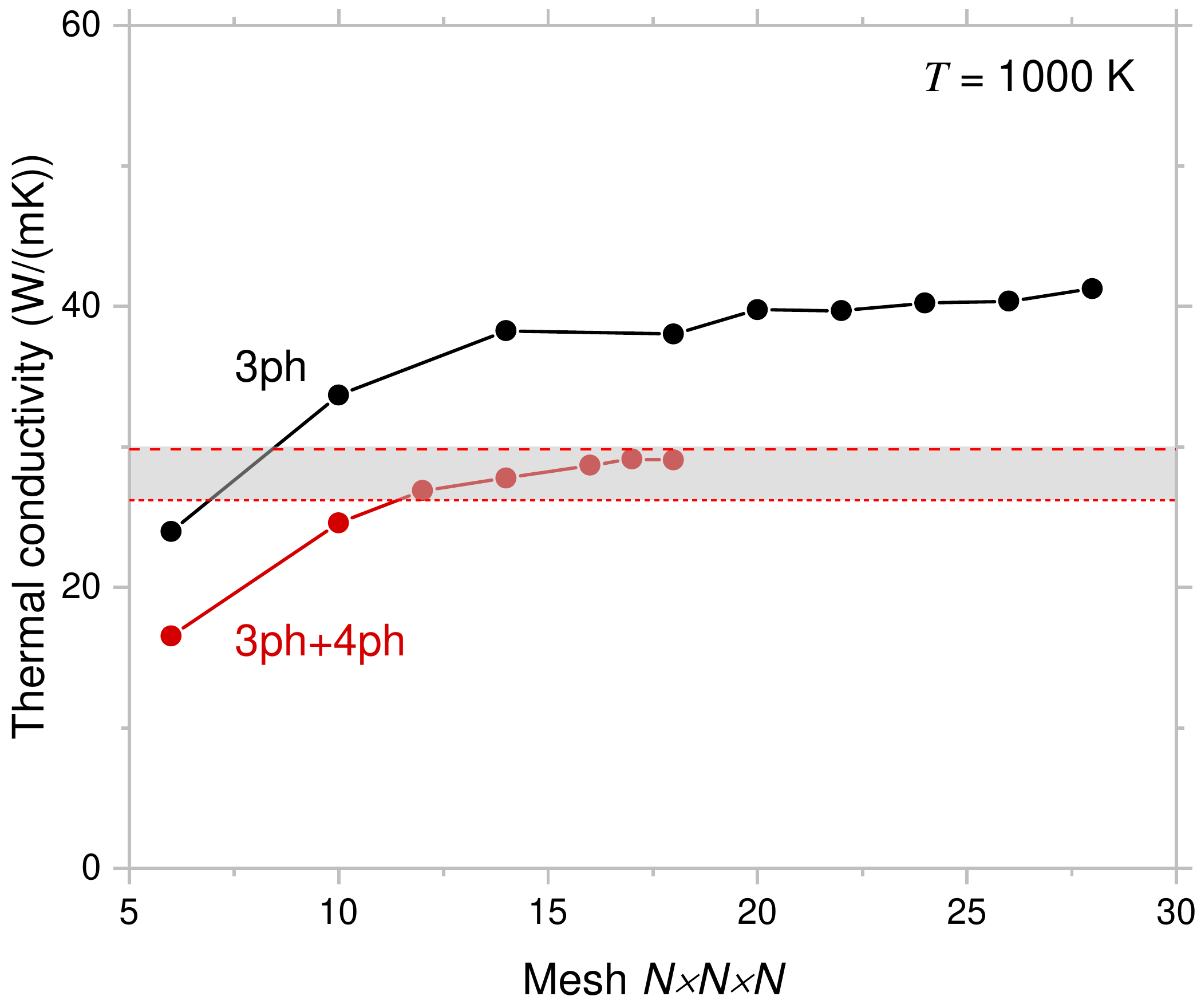}}
\caption{Thermal conductivity of Si at 1000 K calculated by \texttt{ShengBTE} with \texttt{FourPhonon} using first principles. The plot shows the convergence curve with respect of the q-points mesh. The value of \textit{scalebroad} is taken as $0.1$. Experimental data are represented by red straight lines (dot line \cite{glassbrenner1964thermal}, dash line \cite{shanks1963thermal}
).}
\label{Si-converge}
\end{figure}

Regarding the convergence with respect to the $\Vec{q}$-points, our tests on silicon show that (see Fig. \ref{Si-converge}) the inclusion of four-phonon scattering would only require a $17\times17\times17$ $\Vec{q}$-points grid to produce converged thermal conductivity values. In contrast, the three-phonon scattering typically requires a $25\times25\times25$ $\Vec{q}$-points grid to converge. This is beneficial given 
the fact that the calculation of four-phonon scattering is very expensive. 

\begin{figure}
\centerline{\includegraphics[width=6in]{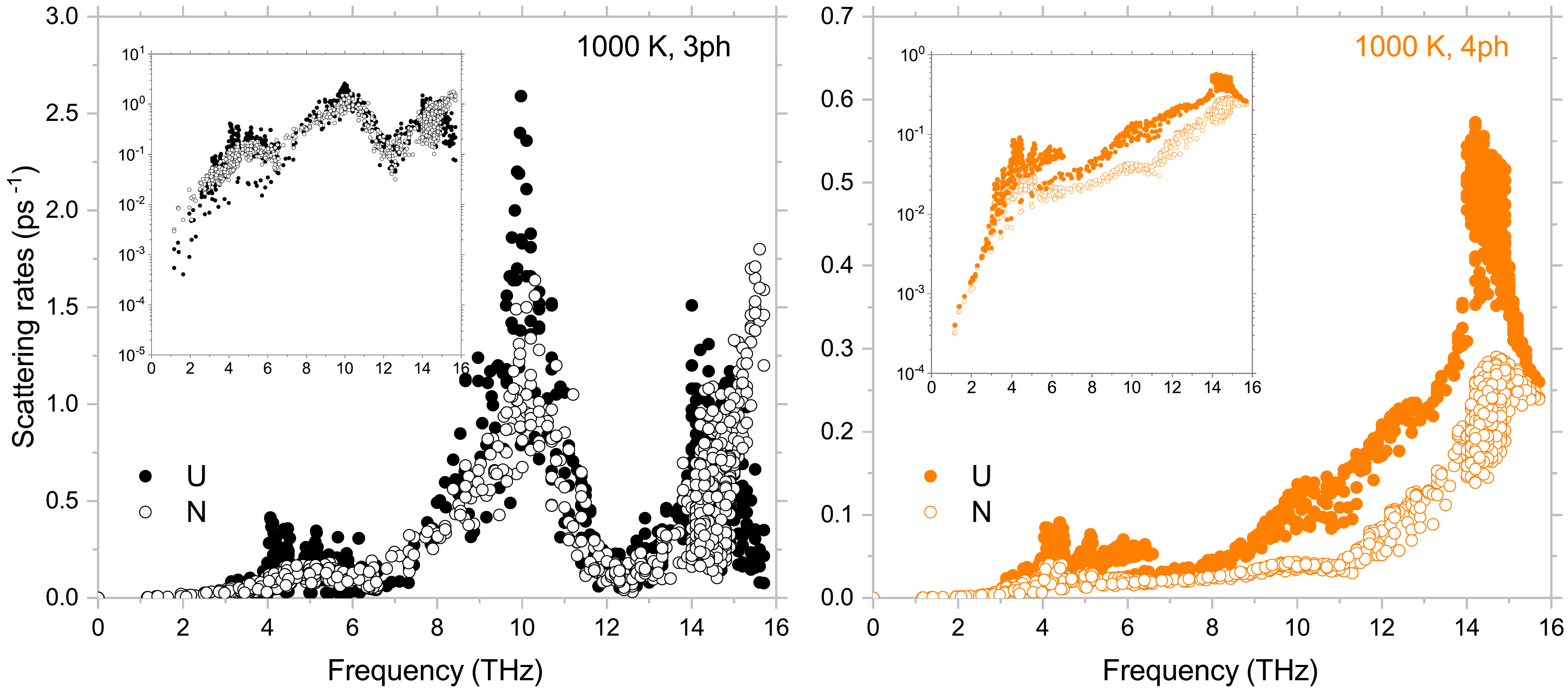}}
\caption{Phonon scattering rates of Si at 1000 K (left: three-phonon, right: four-phonon). The value of \textit{scalebroad} is taken as $0.1$. The inset is log-scale in scattering rates. In both figures, solid circles indicate phonon Umklapp process while hollow ones represent normal process.}
\label{Si-scatter}
\end{figure}

Details on phonon-phonon interactions in Si at room temperature are shown in Fig. \ref{Si-scatter}. Consistent with the literature \cite{prb2017RC}, normal three-phonon processes do not overwhelm Umklapp processes. And in the four-phonon scattering, Umklapp process is dominant. Thus, applying only RTA scheme to Si is acceptable in calculating thermal conductivity.

\subsection{BAs}
Four-phonon effect is known to be important in boron arsenide (BAs), a zinc-blende structure material promising for thermal management applications. When determining the intrinsic thermal transport process, four-phonon scattering is even stronger than three-phonon scatterings. It was found that the higher-order anharmonicity would result in a substantial reduction of thermal conductivity \cite{prb2017RC}, and subsequent experimental studies have validated the predicted thermal conductivity values. Therefore, we use BAs as an example material to further illustrate \verb$FourPhonon$ workflow in calculating thermal conductivity.

\begin{figure}
\centerline{\includegraphics[width=6in]{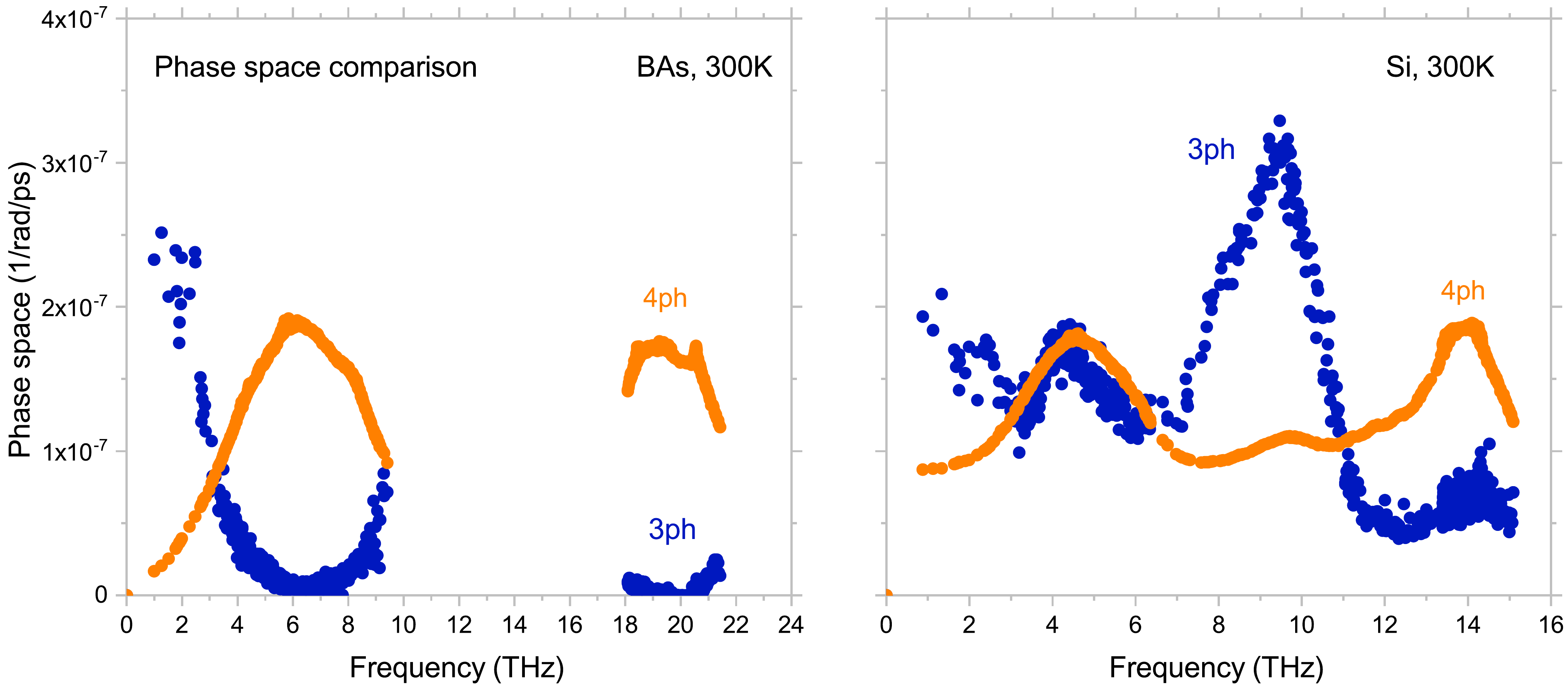}}
\caption{Phase space available of BAs (left) and Si (right) at 300 K. The value of \textit{scalebroad} is taken as $0.1$ and data is presented in log-scale. In both figures, blue circles represent the phase space for three-phonon scattering events ($P_3$), and orange circles represent the phase space for four-phonon scattering events ($P_4$).}
\label{Phasespace}
\end{figure}

The sets of IFCs are obtained from first-principles on a $4\times4\times4$ supercell. Following parameters used in literature, we firstly perform calculations on a $16\times16\times16$ $\Vec{q}$-points grid with \verb$scalebroad$=0.1 at 300 K. For illustrating purpose, we assume that at this stage users have no knowledge on how important four-phonon effect is in BAs. Instead of directly calculating scattering rates with potentially very high cost, one can first look at the weighted phase space as defined in Ref.\cite{2015weightedphasespace}. This is done by turning on \verb$onlyharmonic$ and \verb$four_phonon$ flags in the \verb$CONTROL$ file. Phase space calculation is less expensive to perform since it does not involve anharmonic IFCs. The results are shown in Fig. \ref{Phasespace}. We also plot the results of Si for comparison. It is found that phase space of four-phonon scattering is quite large in BAs, especially in optical branches. In contrast in Si, four-phonon phase space is generally smaller than three-phonon phase space. This is consistent with the fact that four-phonon scattering is not significant in silicon at room temperature and only has minor effect in its thermal conductivity \cite{prb2017RC}. Thus, the calculation of phase space can provide a preliminary estimation if scattering needs to be considered for thermal transport study or not.

\begin{figure}
\centerline{\includegraphics[width=6in]{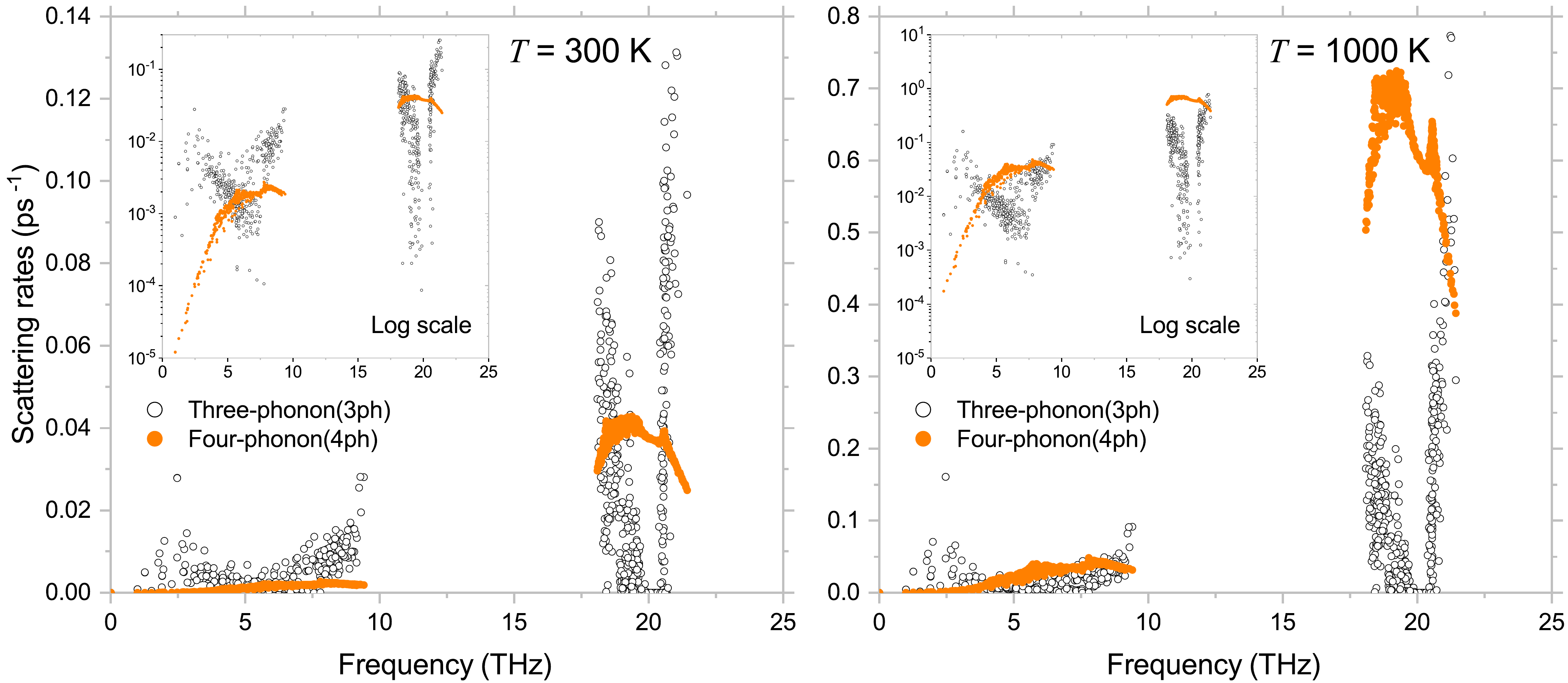}}
\caption{Phonon scattering rates of BAs at 300 K (left) and 1000 K (right). The value of \textit{scalebroad} is taken as $0.1$. The inset is log-scale in scattering rates. One can observe from the logarithm scale that when frequency approaches zero, the scattering rate also goes down to zero, or $\lim\limits_{\omega \to 0} \tau^{-1}=0$.}
\label{BAs-scatter}
\end{figure}

Figure \ref{BAs-scatter} presents the phonon scattering rates in BAs at 300 K and 1000 K. The four-phonon scattering rates can be comparable to or even larger than three-phonon ones even at room temperature. The scattering rates calculation takes about 1680 CPU hours for a single temperature point.

Thermal conductivity is then obtained by exactly solving BTE in an iterative scheme involving three-phonon interactions. To ensure the accuracy of \verb$scalebroad$, we also perform calculations on the same mesh with \verb$scalebroad$=1. The thermal conductivity values in W/mK are 1221.9 (\verb$scalebroad$=1) vs. 1266.9 (\verb$scalebroad$=0.1), and the relative difference is within 4\%. Considering the fact that four-phonon scattering calculations are very expensive at a larger broadening factor, we set \verb$scalebroad$=0.1 in calculations at all temperature points.

\begin{figure}
\centerline{\includegraphics[width=4in]{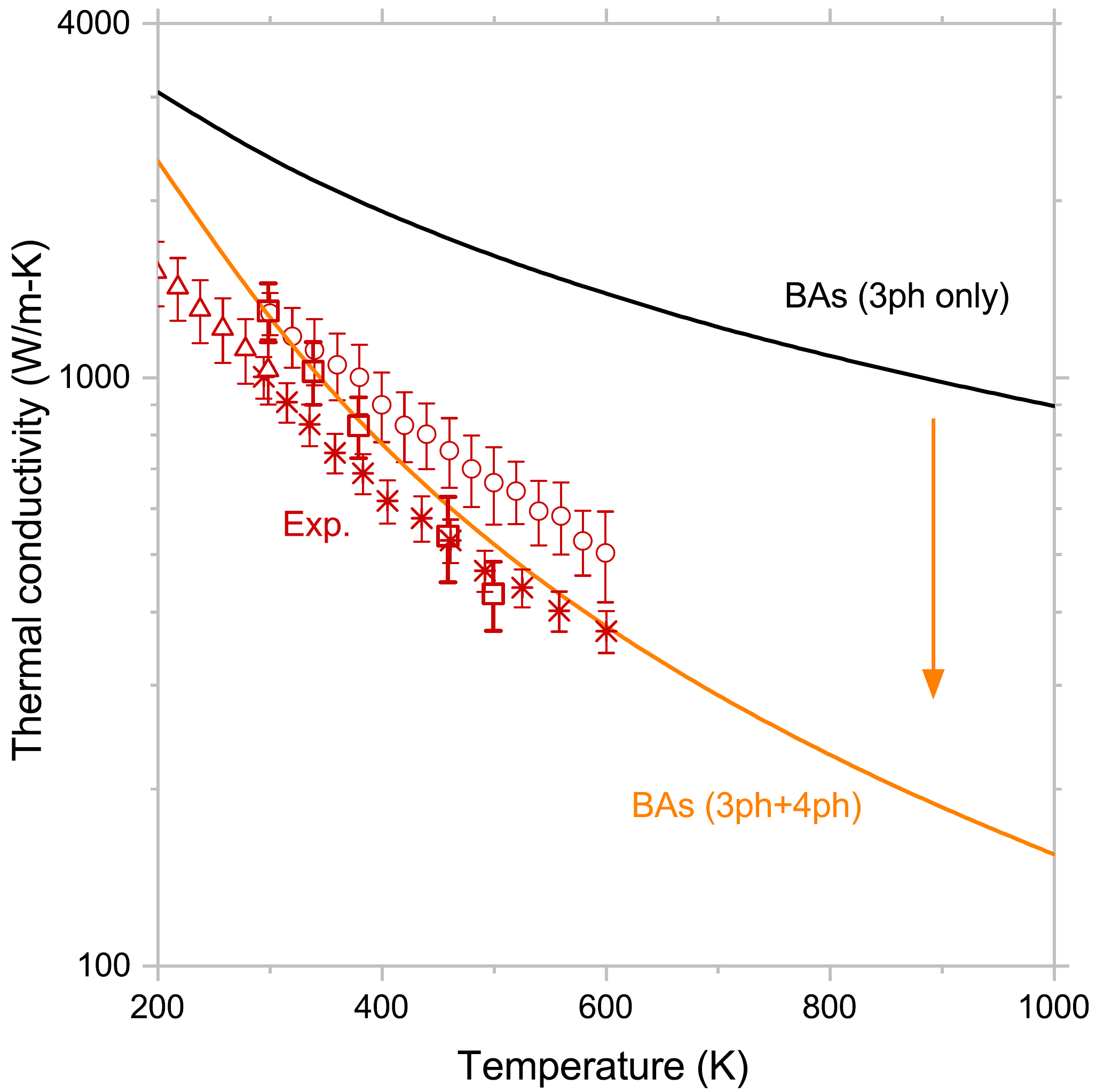}}
\caption{Thermal conductivity of naturally occurring BAs. The value of \textit{scalebroad} is taken as $0.1$. Red symbols are from three experimental works. Red triangle and red square are from Ref.\cite{scienceChen2018}, red circles Ref.\cite{scienceHu2018}, red stars Ref.\cite{scienceCahill2018}.}
\label{BAs-TC}
\end{figure}

After performing calculations from 200 K to 1000 K, we get the temperature-dependent thermal conductivity, as shown in Fig. \ref{BAs-TC}. Consistent with results in Fig. \ref{BAs-scatter}, larger reductions are observed for higher temperatures when considering four-phonon scatterings. Our results are in good agreement with experimental data.

\subsection{$LiCoO_2$}
$\mathrm{LiCoO_2}$ is a representative material of layered lithium transition-metal oxides ($\mathrm{Li_xTMO_2}$). This family of materials store lithium ions in between the $\mathrm{TMO_2}$ layers, and have significant applications in rechargeable batteries due to their supreme electrical and mechanical properties. However, their thermal transport properties seem to be hindering the improvement of device performance \cite{LiBatteryScienAdv,feng2020quantum}. As revealed by Feng et al. \cite{feng2020quantum}, the theoretical upper limit of $\mathrm{LiCoO_2}$'s lattice thermal conductivity would be lowered 2-6 times when including higher-order lattice anharmonicity. Compared to the preceding examples,  lithium  transition-metal oxides have more complex structures and involve four atoms in the unit cell. In this part, we intend to calculate the thermal transport properties of $\mathrm{LiCoO_2}$ using our new computational tool, and decompose the scattering rates into different scattering channels (i.e., recombination, the redistribution and splitting processes).

A $3\times3\times3$ supercell is employed to obtain IFCs by first-principles (for detailed information see Ref. \cite{feng2020quantum}) and the Brillouin zone is discretized by $10\times10\times10$  $\Vec{q}$-points grid. For this material with complex lattice structures, we only use \verb$scalebroad$=0.01 to perform four-phonon calculations, and the available four-phonon processes are already around $10^{10}$.
\begin{figure}
\centerline{\includegraphics[width=6in]{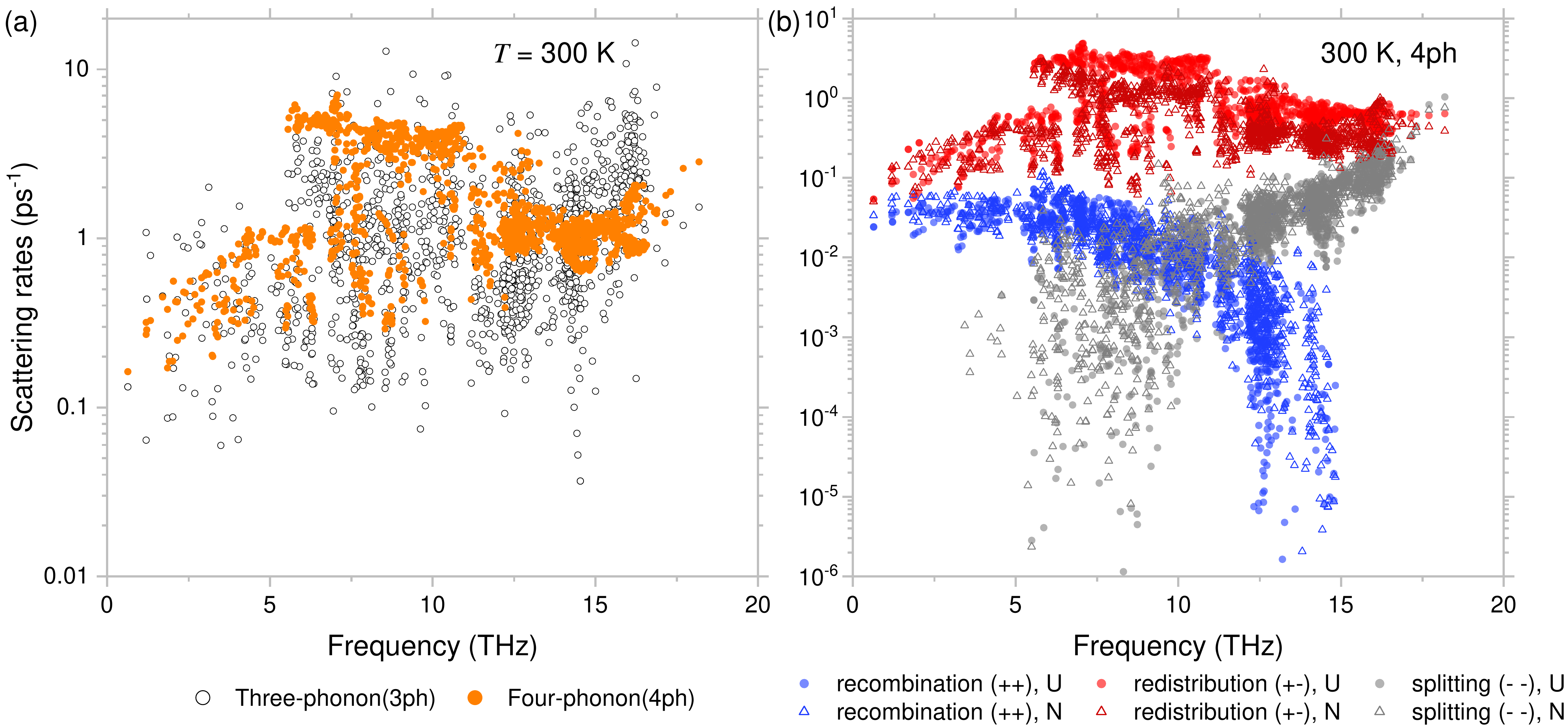}}
\caption{Phonon scattering rates of $\mathrm{LiCoO_2}$ at 300 K. In both figures, scalebroad = 0.01 and y-axis is presented in logarithm scale. Figure (a) shows the comparative strength of three- and four-phonon scattering rates. Captions in (b) represent scattering rates from three scattering channels: recombination ($\lambda+\lambda'+\lambda''\to \lambda'''$), redistribution ($\lambda+\lambda'\to \lambda''+\lambda'''$) and splitting ($\lambda\to\lambda'+\lambda''+\lambda'''$).}
\label{Li-scatter}
\end{figure}

Results of phonon scattering rates are shown in Fig .\ref{Li-scatter}. Even at room temperature,  four-phonon scattering rates of $\mathrm{LiCoO_2}$ are generally comparable to three-phonon scattering rates over the entire frequency range. Different from BAs, four-phonon scattering rates are not that significant in optical phonon branches. Decomposing the four-phonon interactions further into different scattering channels provides us with more information. One can observe that redistribution process($\lambda+\lambda'\to \lambda''+\lambda'''$) contributes the most and Umklapp process dominates this specific scattering channel.

We then calculate the in-plane ($\kappa_\parallel$) and through-plane ($\kappa_\perp$) thermal conductivity and compare their values with literature. Such information is available in a file named \verb$BTE.kappa_tensor$. At 300 K, our calculations yield 9.35 W/mK for $\kappa_\parallel$ and 1.39 W/mK for $\kappa_\perp$, which agree reasonably well with Ref.\cite{feng2020quantum} that gives 9.7 W/mK and 1.4 W/mK, respectively. 

\section{Conclusion}

The recent introduction and implementation of Feng and Ruan's four-phonon calculation framework has led the community to revise the understanding of thermal transport theory. This motivated us to develop a computational package, \verb$FourPhonon$, that can make this calculation accessible to the thermal science community and make researchers able to explore four-phonon effects in a broader range of topics. This new program is developed under a well-perceived  phonon calculation platform \verb$ShengBTE$. Our main contribution is:
\begin{itemize}
    \item Provide the first open-source tool to perform four-phonon scattering calculations, while preserve nice features of \verb$ShengBTE$ platform;
    \item Extend the adaptive broadening scheme to four-phonon scattering cases, provide a script for fourth-order force-constants calculations that utilizes both the point-group and translational symmetries.
\end{itemize}
Examples on silicon, BAs and $\mathrm{LiCoO_2}$ cover common materials, typical materials known for significant four-phonon effects, and newly reported complex systems. These examples show different aspects of our new program and some technical features like convergence with respect to broadening factor and $\Vec{q}$-points grid, scattering contributions from Umklapp and normal processes or from different scattering channels.

Our program is currently able to perform four-phonon calculations under Single Mode Relaxation Time Approximation and iterate with three-phonon scattering rates. We look forward to provide the community with a version capable of fully four-phonon iterations calculations when the memory issue is resolved.

\clearpage

\section*{Acknowledgements}

X. R. and Z. H. acknowledge the partial support from the National Science Foundation (Grant No. 2015946). Simulations were performed at the Rosen Center for Advanced Computing (RCAC) of Purdue University. X.Y. acknowledges support from the Natural Science Foundation of China (Grant No. 12004254). W. L. acknowledges support from the Natural Science
Foundation of China (Grant No. 11704258). 

\appendix
\section{Input and Output Files of \texttt{FourPhonon}}
\label{appendix.files}
Apart from the input files required by \verb$ShengBTE$, \verb$FourPhonon$ takes an additional force constants file, namely \verb$FORCE_CONSTANTS_4TH$. This file can be generated using our script \verb$Fourthorder.py$. Consistent with \verb$ShengBTE$, our program takes no command-line arguments and users only need to prepare a \verb$CONTROL$ file that contains all the settings on crystal structures, energy broadening factor, and $\Vec{q}$-points grid. To enable the four-phonon calculations, users should set \verb$four_phonon=.TRUE.$ in this file. After the computations, we will have these output files besides normal outputs from \verb$ShengBTE$:\\

\noindent\begin{tabularx}{\textwidth}{@{}l|X}
\noindent \verb$BTE.Numprocess_4ph$: & number of allowed four-phonon scattering processes for each mode.\\
\noindent \verb$BTE.P4$: & phase space available for four-phonon scatterings.\\
\noindent \verb$BTE.WP4$: & weighted phase space available for four-phonon processes.\\
\noindent \verb$BTE.w_4ph$: & four-phonon scattering rates under RTA.\\
\noindent \verb$BTE.w_4ph_Umklapp$: & four-phonon scattering rates from Umklapp processes under RTA.\\
\noindent \verb$BTE.w_4ph_normal$: & four-phonon scattering rates from normal processes under RTA.\\
\end{tabularx}\\

\section{Derivation of Adaptive Broadening Formulas}
\label{appendix.formulas}
For a certain phonon mode ($\Vec{q},\omega$), the other three phonons involved are located $\Vec{q}',\Vec{q}'',\Vec{q}'''$ in k-space and their corresponding frequencies are $\omega',\omega'',\omega'''$. The fourth phonon ($\Vec{q}''',\omega'''$) is dependent on the choice of the second and the third phonon while the second phonon ($\Vec{q}',\omega'$) and the third phonon ($\Vec{q}'',\omega''$) are independent from each other. Energy spacing level can be expressed for different scattering processes and then take the derivative:

For recombination (++) process, $\Delta\omega=-\omega'-\omega''+\omega'''$ and $\Vec{q}+\Vec{q}'+\Vec{q}''=\Vec{q}'''$, we have
\begin{align}
    \frac{\partial \Delta\omega}{\partial \vec{q}''}=-\frac{\partial \omega''}{\partial \vec{q}''}+\frac{\partial \omega'''}{\partial \vec{q}'''}\frac{\partial \vec{q}'''}{\partial \vec{q}''}=-\vec{v}_{\lambda''}+\vec{v}_{\lambda'''}
\end{align}
for redistribution (+ -) process, $\Delta\omega=-\omega'+\omega''+\omega'''$ and $\Vec{q}+\Vec{q}'-\Vec{q}''=\Vec{q}'''$, we have
\begin{align}
    \frac{\partial \Delta\omega}{\partial \vec{q}''}=\frac{\partial \omega''}{\partial \vec{q}''}+\frac{\partial \omega'''}{\partial \vec{q}'''}\frac{\partial \vec{q}'''}{\partial \vec{q}''}=\vec{v}_{\lambda''}+\vec{v}_{\lambda'''}(-1)
\end{align}
for splitting (- -) process, $\Delta\omega=\omega'+\omega''+\omega'''$ and $\Vec{q}-\Vec{q}'-\Vec{q}''=\Vec{q}'''$, we have
\begin{align}
    \frac{\partial \Delta\omega}{\partial \vec{q}''}=\frac{\partial \omega''}{\partial \vec{q}''}+\frac{\partial \omega'''}{\partial \vec{q}'''}\frac{\partial \vec{q}'''}{\partial \vec{q}''}=\vec{v}_{\lambda''}+\vec{v}_{\lambda'''}(-1)
\end{align}
energy broadening factor $\sigma=|\frac{\partial \Delta\omega}{\partial \vec{q}''}||\Delta \Vec{q}''|$ then yields
\begin{align}
    \sigma=\begin{cases}
    |-\vec{v}_{\lambda''}+\vec{v}_{\lambda'''}||\Delta \Vec{q}''|,&\text{for recombination (++) process}\\
    |\vec{v}_{\lambda''}-\vec{v}_{\lambda'''}||\Delta \Vec{q}''|,&\text{for redistribution (+ -) process}\\
    |\vec{v}_{\lambda''}-\vec{v}_{\lambda'''}||\Delta \Vec{q}''|,&\text{for splitting (- -) process}.
    \end{cases}
\end{align}
Since we are taking the absolute value, the final formula for all the processes is then
\begin{align}
    \sigma=
    |\vec{v}_{\lambda''}-\vec{v}_{\lambda'''}||\Delta \Vec{q}''|
\end{align}




\bibliographystyle{elsarticle-num.bst}
\bibliography{reference.bib}

\end{document}